\documentclass[12pt]{article}
\begin{document}
\baselineskip=0.20in
\vspace{20mm}
\baselineskip=0.30in
\begin{center}

{\large \bf Spectroscopic Studies of Some Diatomic Molecules using Spectrum Generating Algebra Approach}
\vspace{10mm}

{ \large Oyewumi, K.\,J.\footnote{TWAS-UNESCO Associate, Permanent Address: Theoretical Physics Section, Department of Physics, University of Ilorin, P. M. B. 1515, Ilorin, Kwara state, Nigeria. kjoyewumi66@unilorin.edu.ng} and Sen, K.\,D. \\
School of Chemistry,  University of Hyderabad, \\Hyderabad 50046, India.}
 \vspace{5mm}

\baselineskip=0.20in

\vspace{4mm}

\end{center}

\noindent
\begin{abstract}
For arbitrary values $n$ and $\ell$ quantum numbers, we present
the solutions of the $3$-dimensional Schr\"{o}dinger wave equation
with the pseudoharmonic potential via $SU(1,1)$ Spectrum
Generating Algebra (SGA) approach. The explicit bound state
energies and eigenfunctions are obtained. The matrix elements
$r^{2}$ and $r\frac{d}{dr}$ are obtained (in a closed form)
directly from the creation and annihilation operators. In
addition, the expectation values of $r^{2}$ and $p^{2}$ and the
Heisenberg Uncertainty Products (HUP) for set of diatomic
molecules ($O_{2}, I_{2}, N_{2}, H_{2}$, CO, NO, HCl, CH, LiH,
ScH, TiH, VH, CrH, MnH, TiC, NiC, ScN, ScF, $Ar_{2}$) for arbitrary values
of $n$ and $\ell$ quantum numbers are obtained. The results
obtained are in excellent agreement with the available results in
the literature. It is also shown that the HUP is obeyed for all
diatomic molecules considered.
\end{abstract}


\baselineskip=0.28in \vspace{2mm}
 \vspace{2mm}

\section{Introduction}
The Spectrum Generating Algebra (SGA) methods have been playing an
important role in solving some quantum mechanical problems, since its introduction by
Schr\"{o}dinger, Infeld, Infeld and Hull \cite{Sch4041,Inf4151}.
This technique serves as a useful tool in various fields of physics,
ranging from quantun mechanics (relativistic and nonrelativistic),
mathematical physics, optics, solid state physics, nuclear physics
to chemical physics. This can be achieved through the construction of the
ladder operators (creation and annihilation operators) or raising and lowering operators.
From these ladder operators, the compact dynamical algebraic groups (a suitable Lie algebra)
that such system belongs can be easily realized [3 - 23].

It should be noted that, Schr\"{o}dinger factorization method has
been less frequently applied to physical systems than Infeld-Hull
factorization method, as it has been analyzed in detail by
Mart$\acute{i}$nez et al. \cite{MaE10}, and exemplified later with
a typical system \cite{MaM08}. Mart$\acute{i}$nez and Mota
presented a systematic procedure  of using the factorization
method to construct the generators for hidden and dynamical
symmetries, and applied this study to $2D$ problems of hydrogen
atom, the isotropic harmonic and other radial potential of
interest.

This algebraic approach has been successfully applied to a set of model potentials such as the  with Morse potential, P\"{o}schl-Teller potential, pseudoharmonic potential, infinite square well in $3D$ as well as $N$-dimensions [4, 7 - 23] and their energy spectrum and the eigenfunctions have been studied.

The $SU(1,1)$ dynamical algebra from  the Schr\"{o}dinger ladder operators for hydrogen atom,
Mie-type potential, harmonic oscillator and pseudoharmonic oscillator for $N$-dimensional systems
 have been extensively discussed by Mart$\acute{i}$nez et al. \cite{MaE10}. In the similar fashion,
 Salazar-Ram$\acute{i}$rez et al. \cite{SaE101,SaE102} have applied the factorization method to
 construct the generators of the dynamical algebra $SU(1,1)$ for the
radial equation of the non-relativistic and relativistic generalized MICZ-Kepler problem. It should be noted that
 the generators in the examples\cite{MaM08,MaE10,SaE101, SaE102} above have been constructed
 without adding phase as an additional variable like in Mart$\acute{i}$nez-y-Romero et al. \cite{MaE07}.

Gur and Mann \cite{GuM05} have used the $SU(1,1)$ SGA method to construct the associated
radial Barut-Girardello coherent states for the isotropic harmonic oscillator in arbitrary
dimension and these states have been mapped  into the Sturm-Coulomb radial coherent states.
The dynamics of the $SU(1,1)$ coherent states for the time-dependent quadratic hamiltonian
system has beeen discussed by Choi \cite{Cho09}.

In their work, Motavalli and Akbarieh \cite{MoA10} presented a general construction for the
ladder operators for special orthogonal functions based on the Nikiforov-Uvarov formalism
and generated a list of creation and annihilation operators for some well known special functions.

In the present study, we have followed the approach introduced by 
Dong \cite{Don07}. This is done by using the recursion relations
for the generalized Laguerre polynomials and the explict form of
the eigenfunctions, the $SU(1,1)$ dynamical algebra generators for
some quantum mechanical systems can therefore be obtained.

Apart from generating the eigenvalues and eigenfunctions, this
approach offers the additional advantages in that it can be used
to find the matrix elements in a simple way, and it is also very
useful in constructing coherent states of a given Hamiltonian
system \cite{GuM05,Cho09,DeF11}. Thus, Gur and Mann \cite{GuM05}
used SGA approach to construct the radial Barut-Girardello
 coherent states for the isotropic harmonic oscillator in arbitrary dimension  and mapped these
 states into Sturm-Coulomb radial coherent state; the dynamics of SU(1, 1) coherent states are
 investigated for the time-dependent quadratic Hamiltonian system by Choi \cite{Cho09}.
 Very recently, the second lowest and second highest bases of the discrete positive and
 negative irreducible representations of $SU(1, 1)$ Lie algebra via spherical harmonics
 are used to construct generalized coherent states by Dehghani and  Fakhri \cite{DeF11}.

In recent years,  discussion on the $3D$-dimensional anharmonic
oscillators has been receiving considerable attention in chemical
physics. This is due to their usefulness in studying the dynamical
variables of diatomic molecules. The Morse potential has been one
of the most popular model potential which is employed in the study
of molecular spectra [36 - 38, 41 - 44].

The corresponding wavefunction does not vanish at the origin, and
the exact solutions for any angular momentum ($\ell \neq0$ ) are
as yet unknown. Several other potentials are been used as
alternatives and their performances have been compared with the Morse
potential \cite{QiD07,Haj00, BeS092}. For examples, Kratzer and
pseudoharmonic potentials which have known exact solutions like in
the Coulomb and harmonic oscillator model potentials \cite{Oye10,
SeE071, BeH05,BeE06,SeE072, BaE07, IkS09}.

For the purpose of this study, we consider pseudoharmonic potential.
This potential has been very useful in the area of physical sciences
and it has been extensively used to describe interaction of some
diatomic molecules since its introduction \cite{Don07, OyE0808, GoK61,Sag84, SaG85, OyB03,PaS07,IkS07, SeE071}.
 Sage \cite{Sag84} has discussed the energy levels and wavefunctions of a
 rotating diatomic molecule using a three-parameter model potential called
 the pseudogaussian (pseudoharmonic) potential and he found that the potential
 is reasonably behaved for both small and large internuclear separations.

Obviously, the pseudoharmonic oscillator behaves asymptotically as a
harmonic oscillator, but has a minimum at $r = r_{e}$ and exhibits a
repulsive inverse-square-type singularity at $r=0$.  The energy eigenvalues
and the eigenfunctions of the pseudoharmonic oscillator can be found exactly
for any angular momentum. These wavefunctions have reasonable behaviour at the origin,
near the equilibrium, and at the infinity \cite{ErS88}.

Its characteristics make it useful to model various physical
systems, including some molecular physical ones \cite{ Don07,
Oye10,Sag84,SaG85,IkS07, SeE072}. From the mathematical point of
view, it resembles the harmonic oscillator, from which it deviates
by two correction terms depending on the potential depth and the
equilibrium distance paramater $r_{e}$: the first one is an energy
shift and the second one is a modified centrifugal term. The
latter can also be viewed as originating formally from a
non-integer orbital angular momentum \cite{SaG85}. The
eigenfunctions and  energy eigenvalues are similar to those of the
harmonic oscillator, which can be obtained exactly in the
$r_{e}\rightarrow 0$ limit.

Recently, with an improved approximation to the orbital
centrifugal term of the Manning-Rosen potential, Ikhdair
\cite{Ikh11} used the nikiforov-Uvarov method to obtain the
rotational-vibrational energy states for a few diatomic molecules
for arbitrary quantum numbers $n$ and $\ell$ with different values
of the potential parameters.

In the study of the diatomic molecules using the diatomic
molecular potentials, different methods have been employed:
Nikiforov-Uvarov method [35, 41 - 46]; 
 asymptotic iteration method \cite{BaE07}; Exact
method \cite{IkS07}; shifted $1/N$ expansion \cite{Mor89}; exact
quantization rule method \cite{QiD07, IkS09}; SUSY approach
\cite{Mor04}; Nikiforov-Uvarov method \cite{SeE071, BeH05,
SeE072}; tridiagonal J-matrix representation \cite{NaE0708} and
algebraic method \cite{Oye10, BaE72}.

The aim of this work is to realize the dynamical $SU(1,1)$ algebra
generators for the pseudoharmonic potential to obtain the energy
eigenvalues, eigenfunctions and the matrix elements of the
pseudoharmonic potential. The results obtained are used to
calculate the bound state energies of some neutral diatomic
molecules (metallic hydrides, homogeneous and heterogeneous
diatomic molecules) for any $n$ and $\ell$ quantum numbers.

The scheme of our presentation is as follows: in Section $2$, we
study the $3$-dimensional Schr\"{o}dinger equation for the
pseudoharmonic potential. In Section $3$, we present the formal
solutions of the problem and describe the SGA method used in
constructing the ladder operators for obtaining the energy
eigenvalues, the eigenfunctions and the metrix elements for the
pseudoharmonic potential. We present in Section $4$, the explicit
bound state energies, the numerical values of the expectation
values of $r^{2}$ and $p^{2}$ and the Heisenberg uncertainty
product for the pseudoharmonic potential for the homogeneous
diatomic molecules ($O_{2}, I_{2}, N_{2}, H_{2}$, $Ar_{2}$); the
heterogeneous diatomic molecules (CO, NO, HCl,CH, LiH); the
neutral transition metal hydrides ( ScH, TiH, VH, CrH, MnH); the
transition-metal lithide (CuLi); the transition-metal carbides
(TiC, NiC); the transition-metal nitrite (ScN) and the
transition-metal fluoride(ScF). Finally, in Section $5$, we
discuss our conclusions.

\section{The $3$-dimensional Schr\"{o}dinger equation for the pseudoharmonic potential}
The pseudoharmonic-type potential can be written in the standard form as
 \cite{ Don07, OyE0808,IkS071,IkS072, PaS07,IkS07}
\begin{equation}
V(r) =A r^{2} + \frac{B}{r^{2}} + c.
\label{p1}
\end{equation}
This potential is associated with the following molecular potentials:
\begin{itemize}
    \item Isotropic harmonic oscillator plus inverse quadratic potential
    \begin{equation}
V(r) = \mu \omega^{2} \frac{r^{2}}{2} + \frac{g}{r^{2}},
\label{p2}
\end{equation}
\end{itemize}
here $A = \mu \omega^{2} $, $B = g$ and $c = 0$ \cite{Don07, OyE0808, OyB03, IkS071,IkS072, PaS07}.
\begin{itemize}
\item The pseudoharmonic potential
\begin{equation}
V(r) = D_{e}\left( \frac{r}{r_{e}} - \frac{r_{e}}{r}\right)^{2},
\label{p3}
\end{equation}
\end{itemize}
where $D_{e}$ is the dissociation energy between two atoms in a
solid and $r_{e}$ is the equilibrium intermolecular separation.
Here, we have $A = \frac{D_{e}}{ r_{e}^{2}}$, $B =  D_{e} r_{e}$
and $c =  - 2D_{e} $ [4, 15, 25 - 32, 35].~
The graphs of the potential in equation(\ref{p3}) for some
diatomic molecules is displayed in the figure below {\bf Figure
1.}

\noindent
{\bf Figure 1.} Shapes of the pseudoharmonic potentials for some diatomic molecules.\\

The Schr\"{o}dinger equation for the three-dimension for this potential is
\begin{equation}
\left[ - \frac{\hbar^{2}}{2 \mu}\Delta + \frac{D_{e}}{r_{e}^{2}}r^{2} + \frac{D_{e}r_{e}^{2}}{r^{2}} -2D_{e} \right]\psi(r,\theta, \phi) = E\psi(r,\theta, \phi)
\label{p4}.
\end{equation}
If we propose $\psi_{n, \ell, m}(r, \theta, \phi)$ to have the form
\begin{equation}
\psi_{n, \ell,m}(r,\theta, \phi) = R_{n, \ell,m}(r)Y_{\ell, m}(\theta, \phi)
\label{p5}
\end{equation}
then, equation (\ref{p4}) reduces to two decoupled differential equations, that is, the radial and angular wavefunctions:
\begin{equation}
\left\{ \frac{d ^{2} }{dr ^{2}} + \frac{2}{r} \frac{d }{dr} + \left[ \frac{2 \mu}{\hbar^{2}} \left[E - \left(\frac{D_{e}}{r_{e}^{2}}r^{2} + \frac{D_{e}r_{e}^{2}}{r^{2}} -2 D_{e} \right)  \right] - \frac{\ell~(\ell + 1)}{r^{2}}   \right] \right\} R_{n, \ell}~(r)= 0
\label{p6}
\end{equation}
and
\begin{equation}
L^{2}Y_{\ell, m}(\theta, \phi) = \hbar^{2} \ell~(\ell + 1)Y_{\ell, m}(\theta, \phi),
\label{p7}
\end{equation}
where $\ell = 0, 1, 2, \ldots$ is the orbital angular momentum quantum numbers, $n = 1, 2, 3, \ldots$ is the principal quantum number, $\mu$ is the reduced mass, $\hbar$ is the Planck's constant divided by $2\pi$ and $E$ is the energy eigenvalue. Equation (\ref{p6}) can be rewritten as
\begin{equation}
\left\{ \frac{d ^{2} }{dr ^{2}} + \frac{2}{r} \frac{d }{dr} + \left [ K^{2} + \frac{4 \mu  D_{e}}{\hbar^{2}}  -\frac{2 \mu D_{e}}{\hbar^{2} r_{e}^{2}}r^{2} - \frac{\gamma_{\ell}(\gamma_{\ell} + 1)}{r^{2}} \right ] \right \} R_{n, \ell}~(r)= 0,
\label{p8}
\end{equation}
where
\begin{equation}
\gamma_{\ell} = \frac{1}{2} \left[-1 + \sqrt{(2 \ell + 1)^{2} + \frac{8 \mu D_{e} r_{e}^{2}}{\hbar^{2}}}  \right]
\label{p9}.
\end{equation}

To obtain the relevant algebraic operators for the radial symmetry, equation (\ref{p8}) is solved and the solutions which is a degenerate hypergeometric or Kummer equation (associated Laguerre differential equation) is obtained [52 - 54]. 
Then, the radial funtions  $R_{n, \ell}(r)$ for this potential is obtained as:
\begin{equation}
R_{n, \ell}~(r)  = N_{n,\ell}~r^{ \gamma_{\ell} } e^{- \lambda r^{2}} L_{n}^{ \gamma_{\ell} + \frac{1}{2}}(2 \lambda r^{2}),
\label{p10}
\end{equation}
where
\begin{equation}
\lambda  = \sqrt{\frac{\mu D_{e}}{2 \hbar^{2}r_{e}^{2}}}
\label{p11}.
\end{equation}

$L_{n}^{k}(x)$ is the associated Laguerre functions [52 - 54], 
$N_{n, \ell}$ is the normalization
constant which is determined from the normalization condition
\begin{equation}
\int_{0}^{\infty} R_{n, \ell}~(r) R_{n', \ell}~(r) dr = \delta_{n, n'}
\label{p12}
\end{equation}
as
\begin{equation}
N_{n, \ell}  = \left[ \frac{ 2 (2 \lambda^{2})^{\frac{1}{4}(2
\gamma_{\ell} + 3 )} n!}{ \Gamma{(n + \gamma_{\ell} +
\frac{3}{2})}} \right]^{\frac{1}{2}}. \label{p13}
\end{equation}

\section{The Spectrum Generating Algebra (SGA)}
In a brief introduction, the classical Lie algebra SU(1,1) can be
generated by the elements $K_{0}, K_{1}, K_{2}$  which satisfies
the following commutation relations:
\begin{equation}
[K_{0}, K_{1}] = iK_{2},~~ [K_{1}, K_{2}] = - iK_{0},~~ [K_{2}, K_{0}] = iK_{1}
\label{p14}.
\end{equation}
Alternatively, these can be expressed in terms of the creation and annihilation operators
\begin{equation}
K_{\pm} = K_{1} \pm iK_{2},
\label{p15}
\end{equation}
the commutation relations together with $K_{0}$ can be written as:
\begin{equation}
[K_{0}, K_{\pm}] = \pm K_{\pm},~~ [K_{-}, K_{+}] = 2 K_{0}
\label{p16}.
\end{equation}

Based on the Schr\"{o}dinger factorization method, Infeld-Hull
factorization method, we adopt the factorization method introduced
by Dong \cite{Don07}. This is done by construction of the creation
and annihilation operators through the recursion relations of the
Laguerre functions that evolved, and thereby construct a suitable
Lie algebra in terms of these ladder operators.

In this case, we obtain the differential operators $\hat{\cal{J}}_{\pm}$ with the following property:
\begin{equation}
\hat{\cal{J}}_{\pm}R_{n_{r},\ell}(r) = j_{\pm}R_{n_{r}{\pm}1,\ell}(r),
\label{p17}
\end{equation}
these operators are of the form
\begin{equation}
\hat{\cal{J}}_{\pm} = A_{\pm}(r)\frac{d}{dr} + B_{\pm}(r)
\label{p18}
\end{equation}
and depend only on the physical variable $r$.

On operating the differential operator $\frac{d}{dr}$ on the radial wavefunctions (\ref{p10}), we have,
\begin{equation}
\frac{d }{dr} R_{n,\ell}(r)=\frac{\gamma_{\ell}}{r} R_{n,\ell}(r) -2 \lambda r R_{n,\ell}(r) +  N_{n, \ell}~ r^{\gamma_{\ell}}~\exp (-\lambda ~r^{2} ) \frac{d}{dr}L_{n}^{\gamma_{\ell} + \frac{1}{2}}(2 \lambda ~r^{2} ).
\label{p19}
\end{equation}
In order to find the relationship between $R_{n,\ell}~(r)$ and $R_{n + 1,\ell}~(r)$, the expression above is used to construct the ladder operators $\hat{\cal{J}}_{\pm}$ by using the recurrence relations of the associated Laguerre functions. To find these, the following recurrence relations of the associated Laguerre functions are used [52 - 54]: 
\begin{equation}
x \frac{d}{dx}L_{n}^{\alpha}(x) = \left\{
\begin{array}{ll}
n L_{n}^{\alpha}(x) - (n + \alpha)L_{n-1}^{\alpha}(x) & \\ 
(n + 1) L_{n + 1}^{\alpha}(x) - (n + \alpha + 1 - x)L_{n}^{\alpha}(x)
 & \\  
\end{array}
\right. .
\label{p20}
\end{equation}
The creation and annihilation operators are obtained as:
\begin{equation}
\hat{\cal{J}}_{-}  =\frac{1}{2}\left[- r \frac{d }{dr}  - 2 \lambda r^{2} + 2 \hat {n} + \gamma_{\ell} \right];~~\hat{\cal{J}}_{+}  =\frac{1}{2}\left[ r \frac{d }{dr}  - 2 \lambda r^{2} + 2 \hat {n} + \gamma_{\ell} + 3 \right],
\label{p21}
\end{equation}
where $\hat{n}$ is the number operator with the property
\begin{equation}
\hat{n} R_{n,\ell}(r) = n R_{n,\ell}(r).
\label{p22}
\end{equation}
On defining the operator
\begin{equation}
\hat{\cal{J}}_{0} =  \frac{1}{4} \left[-\frac{d^{2}}{dr^{2}} - \frac{2}{r} \frac{d}{dr} + \frac{\gamma_{\ell}(\gamma_{\ell} + 1 )}{r^{2}}  + \frac{2 \mu D_{e} r^{2}}{\hbar^{2} r_{e}^{2}}  \right],
\label{p23}
\end{equation}
then, the operation of $ \hat{\cal{J}}_{\pm}$ and $\hat{\cal{J}}_{0}$ on the radial wavefunctions $R_{n_{r}, \ell}(r)$ allows us to find the following properties:
\begin{equation}
\hat{\cal{J}}_{+}~R_{n,\ell}(r)  =  \sqrt{(n + 1)\left(n + \gamma_{\ell} + \frac{3}{2}\right)} ~R_{n + 1,\ell}(r) = j_{+}~R_{n + 1,\ell}(r),
\label{p24}
\end{equation}
\begin{equation}
\hat{\cal{J}}_{-}~R_{n,\ell}(r)  =  \sqrt{n \left(n + \gamma_{\ell} + \frac{1}{2}\right)} ~R_{n - 1,\ell}(r) = j_{-}~R_{n - 1,\ell}(r),
\label{p25}
\end{equation}
\begin{equation}
\hat{\cal{J}}_{0}~R_{n,\ell}(r)  = n + \frac{2 \gamma_{\ell} + 3}{4} ~R_{n,\ell}(r) = j_{0}~R_{n,\ell}(r).
\label{p26}
\end{equation}

On carefully inspecting the dynamical group associated to the annihilation and creation operators $ \hat{\cal{J}}_{-} $ and $ \hat{\cal{J}}_{+} $, based on the results of equations (\ref{p24}) and (\ref{p25}), the commutator $[\hat{\cal{J}}_{-},~\hat{\cal{J}}_{+}]$ can be evaluated as follows:\\
\begin{equation}
[\hat{\cal{J}}_{-},~\hat{\cal{J}}_{+} ]R_{n, \ell}(r)
=2 j_{0}R_{n, \ell}(r),
\label{p27}
\end{equation}
where $ j_{0}= n + \frac{2 \gamma_{\ell} + 3}{4}$ and operators $\hat{\cal{J}}_{\mp}$ and $\hat{\cal{J}}_{0}$ satisfy the following commutation relations:
\begin{equation}
[\hat{\cal{J}}_{0},~\hat{\cal{J}}_{\mp} ]R_{n, \ell}(r)
=\mp\hat{\cal{J}}_{\mp} R_{n, \ell}(r).
\label{p28}
\end{equation}

 We define the Hermitian operators for these operators as follows:
\begin{equation}
\displaystyle{\hat{\cal{J}}_{x} = \frac{1}{2} \left( \hat{\cal{J}}_{+} + \hat{\cal{J}}_{-} \right)},~~~
\displaystyle{\hat{\cal{J}}_{y} = \frac{1}{2 i} \left( \hat{\cal{J}}_{+} - \hat{\cal{J}}_{-} \right)},~~~
\displaystyle{\hat{\cal{J}}_{z} = \hat{\cal{J}}_{0}}
\label{p29}
\end{equation}
and the following commutation relations are obtained:
\begin{equation}
\displaystyle{[\hat{\cal{J}}_{x},~\hat{\cal{J}}_{y} ] = - i \hat{\cal{J}}_{z}},~~
\displaystyle{[\hat{\cal{J}}_{y},~\hat{\cal{J}}_{z} ] = i \hat{\cal{J}}_{x}},~~
\displaystyle{[\hat{\cal{J}}_{z},~\hat{\cal{J}}_{x} ] = i \hat{\cal{J}}_{y}}.
\label{p30}
\end{equation}

The Casimir operator can be expressed as \cite{Cas31}
\begin{equation}
\begin{array}{rl}
&\hat{\cal{C}}R_{n, \ell}(r)= \left \{ \hat{\cal{J}}_{0}(\hat{\cal{J}}_{0} - 1 ) - \hat{\cal{J}}_{+} \hat{\cal{J}}_{-}\right \} R_{n, \ell}(r)
=\left \{\hat{\cal{J}}_{0}( \hat{\cal{J}}_{0} + 1) - \hat{\cal{J}}_{-} \hat{\cal{J}}_{+} \right \}R_{n, \ell}(r)
\nonumber\\
& = \left( \frac{2 \gamma_{\ell} + 3}{4} \right) \left ( \frac{2 \gamma_{\ell} - 1}{4} \right )R_{n, \ell}(r) = \tau(\tau - 1)R_{n, \ell}(r),
\label{p31}
\end{array}
\end{equation}
where
\begin{equation}
\tau = \frac{2 \gamma_{\ell} + 3}{4}.
\label{p32}
\end{equation}
Then, the Casimir operator $\hat{\cal{C}} $ now satisfies the following commutation relations:
\begin{equation}
[ \hat{\cal{C}},  \hat{\cal{J}}_{\pm}] =[ \hat{\cal{C}},  \hat{\cal{J}}_{x}] =
[\hat{\cal{C}}, \hat{\cal{J}}_{y}] =  [\hat{\cal{C}}, \hat{\cal{J}}_{z}] = 0,
\label{p33}
\end{equation}
therefore, the operators $\hat{\cal{J}}_{\pm} $, $\hat{\cal{J}}_{x} $, $\hat{\cal{J}}_{y} $, $\hat{\cal{J}}_{z} $ and $\hat{\cal{J}}_{0} $ satisfy the commutation relations of the dynamical group $SU(1,1)$ algebra, which is isomorphic to an $SO(2, 1)$ algebra (i. e. $SU(1,1)$ $\sim$ $SO(2, 1)$. The commutation rules are valid for the infinitesimal operators of the non-compact group $SU(1, 1)$ \cite{BaR00, NiE91}.

The Hamiltonian operator $\hat{\cal{H}}$ takes the form
\begin{equation}
\hat{\cal{H}}= \left(4n + 2\gamma_{\ell} + 3 \right)\sqrt{\frac{ \hbar^{2} D_{e}}{2 \mu r_{e}^{2}}} - 2 D_{e} = \frac{1}{4}\hat{\cal{J}}_{0}\sqrt{\frac{ \hbar^{2} D_{e}}{2 \mu r_{e}^{2}}} - 2D_{e}~ .
\label{p34}
\end{equation}

Furthermore, we find that the following physical functions can be obtained by the creation and annihilation operators $\hat{\cal{J}}_{\mp}$ and $\hat{\cal{J}}_{0}$ as:
\begin{equation}
r^{2}R_{n, \ell}(r)= \frac{1}{2 \lambda}[2 \hat{\cal{J}}_{0} -(\hat{\cal{J}}_{+} + \hat{\cal{J}}_{-})]R_{n, \ell}(r),
\label{p35}
\end{equation}
\begin{equation}
r \frac{d}{dr}R_{n, \ell}(r)=  (\hat{\cal{J}}_{+} - \hat{\cal{J}}_{-}) - \frac{3}{2}R_{n, \ell}(r).
\label{p36}
\end{equation}
With equations (\ref{p35}) and (\ref{p36}), the matrix elements for $r^{2}$ and $r\frac{d}{dr}$ are obtained as follows:
\begin{equation}
\begin{array}{rl}
&\langle R_{m, \ell}(r) \mid r^{2}\mid R_{n, \ell}(r)\rangle
\nonumber \\
&=\displaystyle{\frac{1}{ 2 \lambda}\left[\left(2 n + \frac{2 \gamma_{\ell} + 3}{2}\right) \delta_{m, n} - j_{+} \delta_{m, n + 1}  - j_{-} \delta_{m, n - 1}\right]}
\label{p37}
\end{array}
\end{equation}
and
\begin{equation}
\begin{array}{rl}
&\langle R_{m, \ell}(r) \mid r\frac{d}{dr}\mid R_{n, \ell}(r)\rangle
\nonumber \\
& =\displaystyle{j_{+} \delta_{m, n + 1} - j_{-} \delta_{m, n - 1}  - \frac{3}{2} \delta_{m, n}}.
\label{p38}
\end{array}
\end{equation}

From the relations above, we can deduce the following relations:
\begin{equation}
2 \lambda \langle R_{m, \ell}(r) \mid r^{2}\mid R_{n, \ell}(r)\rangle + \langle R_{m, \ell}(r) \mid r\frac{d}{dr}\mid R_{n, \ell}(r)\rangle 
=(2 n +  \gamma_{\ell} ) \delta_{m, n} - 2 j_{-} \delta_{m, n - 1}
\label{p39}
\end{equation}

and
\begin{equation}
2 \lambda \langle R_{m, \ell}(r) \mid r^{2}\mid R_{n, \ell}(r)\rangle - \langle R_{m, \ell}(r) \mid r\frac{d}{dr}\mid R_{n, \ell}(r)\rangle  
=(2 n +  \gamma_{\ell} + 3 ) \delta_{m, n} - 2 j_{+} \delta_{m, n + 1}.
\label{p40}
\end{equation}
These two relations form a useful link for finding the matrix elements from the creation and annihilator operators.

\section{The explicit bound state energies, the numerical values of $\langle r^{2} \rangle$, $\langle p^{2} \rangle $ and the Heisenberg Uncertinty Products}

\subsection{The explicit bound state energies for the pseudoharmonic potential}

The explicit bound state energies for the pseudoharmonic potential are obtained as :
\begin{equation}
E_{n, \ell} = \left(4n + 2\gamma_{\ell} + 3 \right)\sqrt{\frac{ \hbar^{2} D_{e}}{2 \mu r_{e}^{2}}} - 2 D_{e},
\label{p41}
\end{equation}
where $\gamma_{\ell} $ is as stated in equation (\ref{p9}).

\subsection{The expectation values of $ r^{2}$ and $ p^{2} $ for the pseudoharmonic potential}
The expectation values of $ r^{2}$ and $ p^{2} $ can be obtained by applying the Hellmann-Feynman theorem~(HFT) [30, 57 - 64]. 
This theorem states that a non-degenerate eigenvalue of a hermitian operator in a parameter dependent eigensystem varies with respect to the parameter according to the formula
\begin{equation}
\frac{\partial E_{\nu}}{\partial \nu}=\langle \Psi_{\nu} \mid
\frac{\partial H_{\nu}}{\partial \nu} \mid \Psi_{\nu}\rangle,
\label{p42}
\end{equation}
provided that the associated normalized eigenfunction $\Psi_{\nu}$, is continuous with respect to the
parameter, $\nu$.
The effective Hamiltonian of the pseudoharmonic potential radial wave function is
\begin{equation}
\hat{\cal{H}}= \frac{-\hbar^2}{2 \mu} \frac{d^2}{d r^2}
+\frac{\hbar^2}{2 \mu} \frac{\ell (\ell + 1)}{r^2} +  \frac{D_{e}}{r_{e}^{2}} r^{2} + \frac{D_{e}r_{e}^{2}}{r^{2}} -2 D_{e}
\label{p43}.
\end{equation}
With $\nu = D_{e}$ and $\nu = \mu$, then, the following
expectation values of $r^{2}$ and $p^{2}$ are obtained
respectively as:
\begin{equation}
\langle r^{2} \rangle = \left[2n + 1 + \frac{1}{2} \sqrt{(2 \ell + 1)^{2} + \frac{8 \mu D_{e} r_{e}^{2}}{\hbar^{2}}}   \right]\sqrt{\frac{ \hbar^{2} r_{e}^{2}}{2 \mu D_{e}}}
\label{p44}
\end{equation}
and
\begin{eqnarray}
&\displaystyle{\langle p^{2} \rangle = 2 \mu D_{e}\left[2n + 1 + \frac{1}{2} \sqrt{(2 \ell + 1)^{2} + \frac{8 \mu D_{e} r_{e}^{2}}{\hbar^{2}}}   \right]\sqrt{\frac{ \hbar^{2} }{2 \mu D_{e} r_{e}^{2}}}}\\
\nonumber
& - ~\displaystyle{\frac{ 4 \mu D_{e}  }{\sqrt{(2 \ell + 1)^{2} + \frac{8 \mu D_{e} r_{e}^{2}}{\hbar^{2}}}  }\sqrt{\frac{2 \mu D_{e} r_{e}^{2}}{ \hbar^{2} }}}  .
\label{p45}
\end{eqnarray}

\subsection{The Heisenberg Uncertainty product for the pseudoharmonic potential}
In 1927, Werner Heisenberg stated that certain physical quantities, like the position and momentum, cannot both have precise values at the same time, this is called the Heisenberg uncertainty principle \cite{Hei27}. That is, the more precisely one property is measured, the less precisely the other can be measured. A mathematical statement of this principle is that every quantum state has the property that the root mean square (RMS) deviation of the position from its mean (the standard deviation of the x-distribution):
\begin{equation}
\Delta x =  \sqrt{\langle (x - \langle x \rangle )^{2}    \rangle}
\label{p46}
\end{equation}
 and the RMS deviation of the momentum from its mean (the standard deviation of p):
\begin{equation}
\Delta p_{x} =  \sqrt{\langle (p_{x} - \langle p_{x} \rangle )^{2}    \rangle},
\label{p47}
\end{equation}
the product of which can never be smaller than a fixed fraction of Planck's constant:
\begin{equation}
\Delta x \Delta p_{x} \geq \frac{\hbar}{2}.
\label{p48}
\end{equation}

This inequality is very important in physics, it has been pointed
out that for a particle moving non-relativistically in a central
potential V(r), the following 'uncertainty' relation holds
\cite{PaS07,GrE04,OyE04, Oye06,Hei27,Wok7401}:
\begin{equation}
\Delta r \Delta p \geq  \frac{3 \hbar}{2}. 
\label{p49}
\end{equation}
With equations (\ref{p44}) and (\ref{p45}) and noting that
$\langle r \rangle = \langle p \rangle = 0$ (due to parity
consideration), the Heisenberg uncertainty product for the
pseudoharmonic potential becomes:
\begin{eqnarray}
&\displaystyle{P_{n, \ell} = \Delta r \Delta p = \left[2n + 1 +
\frac{1}{2} \sqrt{(2 \ell + 1)^{2} + \frac{8 \mu D_{e}
r_{e}^{2}}{\hbar^{2}}}   \right]^{2} \hbar^{2}} \\ \nonumber & -
\displaystyle{ \frac{ 4 \mu D_{e} r_{e}^{2}   }{ \sqrt{(2 \ell +
1)^{2} + \frac{8 \mu D_{e} r_{e}^{2}}{\hbar^{2}}} }  \left[2n + 1
+ \frac{1}{2} \sqrt{(2 \ell + 1)^{2} + \frac{8 \mu D_{e}
r_{e}^{2}}{\hbar^{2}}}   \right] }. 
\label{p50}
\end{eqnarray}

\begin{table}[!hbp]
\caption{Model parameters for some diatomic molecules in our study.
\vspace*{13pt}} {\small
\begin{tabular}{c c c c c c c  c c c}
\hline\hline
{}&{}   &{}&{}&{}&{}&{}&{}\\[-1.5ex]
Molecules &$D_{e}$(in eV)  &$r_{e}$(in $A^{o}$)&$\mu$(in amu)& Molecules &$D_{e}$(in eV)  &$r_{e}$(in $A^{o}$)&$\mu$(in
amu)   \\[1ex] \hline\hline
$ScH$  &2.25 &1.776 &0.986040 & $CH$  &3.947418665   &1.1198  &0.929931                 \\[1ex]
$TiH$  &2.05 &1.781 &0.987371  &$LiH$   & 2.5152672118 &1.5956  &0.8801221               \\[1ex]
$VH$   &2.33&1.719 &  0.988005 &$HCl$ &4.619030905  &1.2746 &  0.9801045          \\[1ex]
$CrH$  &2.13 &1.694 &0.988976  &$H_{2}$   &4.7446   &0.7416 &0.50391    \\[1ex]
$MnH$  &1.67 & 1.753 &0.989984  &$CO$  & 10.845073641   &1.1282 &6.860586000        \\[1ex]
$CuLi$ &1.74 &2.310 &6.259494  &$NO$  & 8.043729855   &1.1508 &7.468441000    \\[1ex]
$TiC$  &2.66 &1.790 & 9.606079 &$O_{2}$ &5.156658828  &1.208 &  7.997457504    \\[1ex]
$NiC$  &2.76 &1.621 & 9.974265 &$I_{2}$ &1.581791863  &2.662 &63.45223502    \\[1ex]
$ScN$  &4.56 & 1.768&10.682771 &$N_{2}$ &11.938193820  &1.0940  &7.00335                      \\[1ex]
$ScF$  &5.85 & 1.794&13.358942 &$Ar_{2}$ &1.672 &2.53  &53.9341                    \\[1ex] \hline\hline
\end{tabular}\label{tab1} }
\vspace*{-13pt}
\end{table}

In this work, we obtained the explicit bound state energies ($E_{n, \ell}$), the expectation values of $r^{2}$ and $  p^{2} $ ($\langle r^{2}\rangle$ and $\langle p^{2} \rangle$) and the Heisenberg uncertainty product $ P_{n, \ell} $ of some diatomic molecules for various values of $n$ and $\ell$. In the case of this study, we have selected some diatomic molecules for the purposes which they serve in various aspect of chemical synthesis, nature of bonding, temperature stability and electronic transport properties in chemical physics \cite{BeS091,Har00, Mor86}.

Some of these selected diatomic molecules composed of the some homogeneous diatomic molecules (dimers) ($O_{2}, I_{2}, N_{2}, H_{2}, Ar_{2}$); the heterogeneous diatomic molecules (CO, NO, HCl, CH, LiH); the neutral transition metal hydrides ( ScH, TiH, VH, CrH, MnH); the transition-metal lithide (CuLi); the transition-metal carbides (TiC, NiC); the transition-metal nitrite (ScN) and the transition-metal fluoride(ScF). 

The spectroscopic parameters and reduced masses for some selected diatomic molecules used in our study are shown in Table 1. The spectroscopic parameters listed in this table are obtained from the following cited sources:  for CO, NO, $O_{2}$, $I_{2}$ and $N_{2} $, the sources are  \cite{IkS07,SeE071, BeE06,BaE07,IkS09,KaP70,BrR69}; for CH, LiH and HCl,  the sources are \cite{IkS07,SeE071,QiD07,Ikh09,BeE06,IkS09,KaP70,ViS82} ; for $H_{2}$, the source is \cite{CaE06} ; for $Ar_{2}$, the source is \cite{Var08, Mor86} and  for ScH, TiH, VH, CrH, MnH, CuLi, TiC, NiC, ScN and ScF,  the sources are \cite{BeS091, Har00}, where $\hbar c = 1973.29~ eV A^{o}$ is taken from \cite{IkS07,SeE071, BeS091,Ikh09,BeE06,BaE07,IkS09, Ikh11,NaE10}.

For these selected diatomic molecules, we have available results from the literature to compare with our results with few ones: CO, NO, CH and $ N_{2}$ \cite{IkS07, SeE071}.

The explicit bound state energies of some of these diatomic molecules for various values of $n$ and $\ell$ are obtained and compared with the exact method \cite{IkS07} and the Nikiforov-Uvarov method \cite{SeE071} for CO, NO,CH and $ N_{2}$. In the Tables 2 - 7, we have used $E_{n, \ell}$ (FM) to mean factorization method (present method), $E_{n, \ell}$(EM) to mean Exact Method \cite{IkS07} and $E_{n, \ell}$ (NU) mean Nikiforov-Uvarov Method  \cite{SeE071}.  Also, the numerical results for the expectation values of $r^{2}$ and $  p^{2} $ ($\langle r^{2}\rangle$ and $\langle p^{2} \rangle$) and the Heisenberg uncertainty product $ P_{n, \ell} $ of some of these diatomic molecules for various values of $n$ and $\ell$ are computed and presented in Tables 2 - 7.

\begin{table}[!hbp]
\caption{The energy eigenvalues $E_{n, \ell}$(in eV), the expectation values  $\langle r^{2} \rangle $ (in $ (A^{o})^{2}$), $ \langle p^{2} \rangle $ (in $(eV/c)^{2}$) and the Heisenberg Uncertainty Relations HUR (in $ eV A^{o}/c$) corresponding to the pseudoharmonic potential for various $n$ and $\ell$ quantum numbers for $CO$ and $NO$ diatomic molecules.  \vspace*{13pt}} {\tiny
\begin{tabular}{c c c c c c c c}\hline
{}&{} &{} &{}&{}&{}&{} &{}\\[-1.5ex]
$n$&$\ell$&$E_{n, \ell}$ [FM] & $E_{n, \ell}$ [EM]& $E_{n, \ell}$ [NU]& $\langle r^{2} \rangle$ & $\langle p^{2} \rangle$ & HUR\\[2ex]\hline
\multicolumn{7}{c}{ \bf CO } \\ \hline
$0$&$0$&0.1019578053531127 &0.1019306&0.10193061&1.278818514996878&651953274.6811209&28874.38170030767\\[1ex]
$1$&$0$&0.3057536875651401&0.3056722&0.30567217&1.290777797622587&1954330222.150913&50225.55186332170\\[1ex]
&$1$&0.3062325780354271&0.3061508&0.30615078&1.290805888935957&1960448477.813082&50304.65624687896\\[1ex]
$2$&$0$&0.5095495697771710&0.5094137&0.50941373&1.302737080248297&3256707169.620706&65135.49868831414\\[1ex]
&$1$&0.5100284602474545&0.5098923&0.50989234&1.302765171561666&3262825425.282874&65197.35673280329\\[1ex]
&$2$&0.5109862411880286&0.5108495&0.51084953&1.302821352328619&3275061126.516209&65320.89685396806\\[1ex]
$4$&$0$&0.9171413342012258&0.9168969&0.91689685&1.326655645499716&5861461064.560290&88182.42688980432\\[1ex]
&$1$&0.9176202246715128&0.9173755&0.91737546&1.326683736813085&5867579320.222428&88229.37185880831\\[1ex]
&$2$&0.9185780056120834&0.9183327&0.91833265&1.326739917580037&5879815021.455794&88323.18663268512\\[1ex]
&$3$&0.9200131221837502&0.9197684&0.91976835&1.326824184081614&5898166548.480750&88463.72148098587\\[1ex]
&$4$&0.9219286840648948&0.9216825&0.92168247&1.326936530740917&5922631472.632870&88650.75329151144\\[1ex]
$5$&$0$&1.120937216413253&1.1206384&1.12063840&1.338614928125425&7163838012.030083&97926.60775078365\\[1ex]
&$1$&1.121416106883540&1.1211170&1.12111700&1.338643019438795&7169956267.692221&97969.44374358583\\[1ex]
&$2$&1.122373887824111&1.1220742&1.12207420&1.338699200205747&7182191968.925556&98055.05925001923\\[1ex]
&$3$&1.123809004395778&1.1235099&1.12350990&1.338783466707324&7200543495.950542&98183.34168119120\\[1ex]
&$4$&1.125724566276922&1.1254240&1.12542400&1.338895813366626&7225008420.102662&98354.12307175575\\[1ex]
&$5$&1.128117463789163&1.1278165&1.12781650&1.339036232750661&7255583505.037642&98567.18116591059\\[1ex]
\hline
\multicolumn{7}{c}{ \bf NO } \\ \hline
$0$&$0$&0.08251086588683876 &0.0824883&0.08248827&1.331132968244588&574375573.8664465 &27650.86368502856\\[1ex]
$1$&$0$&0.2474257254889523 &0.2473592&0.24735916&1.344708974060483&1721656608.599298&48115.76760100594\\[1ex]
&$1$&0.2478484808663488 &0.2477817&0.24778171&1.344743765885460&1727536886.807932&48198.49021361709\\[1ex]
$2$&$0$&0.4123405850910657 &0.4122301&0.41223005&1.358284979876379&2868937643.332150&62424.63383264652\\[1ex]
&$1$&0.4127633404684623 &0.4126526&0.41265260&1.358319771701355&2874817921.540784&62489.37527988453\\[1ex]
&$2$&0.4136075930227321 &0.4134977&0.41349768&1.358389352609480&2886577551.185603& 62618.65705212839\\[1ex]
$4$&$0$&0.7421703042952892 &0.7419718&0.74197183&1.385436991508170&5163499712.797879&84579.56909178477\\[1ex]
&$1$&0.7425930596726857 &0.7423944&0.74239438&1.385471783333147&5169379991.006488&84628.77829005003\\[1ex]
&$2$&0.7434373122269555 &0.7432395&0.74323946&1.385541364241271&5181139620.651306&84727.11052739678\\[1ex]
&$3$&0.7447055783591487 &0.7445070&0.74450700&1.385645728749967&5198776748.734846&84874.39423412152\\[1ex]
&$4$&0.7463953416682116 &0.7461969&0.74619689&1.385784868638066&5222288597.129426&85070.37391220915\\[1ex]
$5$&$0$&0.9070851638974027 &0.9068427&0.90684272&1.399012997324066&6310780747.530731&93962.03642460064\\[1ex]
&$1$&0.9075079192747992 &0.9072653&0.90726527&1.399047789149042&6316661025.739365&94006.97124609737\\[1ex]
&$2$&0.9083521718290690 &0.9081104&0.90811035&1.399117370057167&6328420655.384158&94096.77605516852\\[1ex]
&$3$&0.9096204379612622 &0.9093779&0.90937789&1.399221734565862&6346057783.467698&94231.32164752261\\[1ex]
&$4$&0.9113102012703251 &0.9110678&0.91106778&1.399360874453962&6369569631.862278&94410.41536789360\\[1ex]
&$5$&0.9134227199567881 &0.9131799&0.91317990&1.399534778764948&6398952498.950474&94633.80247007970\\[1ex]
\hline\hline
\end{tabular}\label{tab2} }
\vspace*{-13pt}
\end{table}

\begin{table}[!hbp]
\caption{The energy eigenvalues $E_{n, \ell}$(in eV), the expectation values  $\langle r^{2} \rangle $ (in $ (A^{o})^{2}$), $ \langle p^{2} \rangle $ (in $(eV/c)^{2}$) and the Heisenberg Uncertainty Relations HUR (in $ eV A^{o}/c$) corresponding to the pseudoharmonic potential for various $n$ and $\ell$ quantum numbers for $N_{2}$ and $CH$ diatomic molecules.  \vspace*{13pt}} {\tiny
\begin{tabular}{c c c c c c c c}\hline
{}&{} &{} &{}&{}&{}&{} &{}\\[-1.5ex]
$n$&$\ell$&$E_{n, \ell}$ [FM] & $E_{n, \ell}$ [EM]& $E_{n, \ell}$ [NU]& $\langle r^{2} \rangle$ & $\langle p^{2} \rangle$ & HUR\\[2ex]\hline
\multicolumn{7}{c}{ \bf $N_{2}$ } \\ \hline
$0$&$0$&0.1091860343455160&0.1091559&0.10915590&1.202309005240784&712683651.7236753&29272.27309853578\\[1ex]
$1$&$0$&0.3274313753502547&0.3273430&0.32734304&1.213248845205456&2136424222.911053&50911.82791175150\\[1ex]
&$1$&0.3279292342137836&0.3278417&0.32784167&1.213273844148512&2142930999.051733&50989.82379813161\\[1ex]
$2$&$0$&0.5456767163549969& 0.5455302&0.54553018&1.224188685170128&3560164794.098396&66017.52387265288\\[1ex]
&$1$&0.5461745752185259&0.5460288&0.54602881&1.224213684113184&3566671570.239111&66078.49985452284\\[1ex]
&$2$&0.5471719580253911&0.5470260&0.54702603&1.224263680432900&3579684307.137500&66200.28311604151\\[1ex]
$4$&$0$&0.9821673983644743&0.9819045&0.98190446&1.246068365099471&6407645936.473118&89355.27346607656\\[1ex]
&$1$&0.9826652572280032&0.9824031&0.98240309&1.246093364042527&6414152712.613832&89401.52756605153\\[1ex]
&$2$&0.9836626400348720&0.9834003&0.98340031&1.246143360362243&6427165449.512221&89493.96376772718\\[1ex]
&$3$&0.9851595467850807&0.9848961&0.98489606&1.246218350926318&6446682516.785655&89632.43863141369\\[1ex]
&$4$&0.9871526473190038&0.9868903&0.98689026&1.246318331037525&6472701469.817430&89816.73837969733\\[1ex]
$5$&$0$&1.200412739369217&1.2000916&1.20009160&1.257008205064143&7831386507.660496&99217.52414345898\\[1ex]
&$1$&1.200910598232742&1.2005902&1.20059020&1.257033204007199&7837893283.801210&99259.72046708141\\[1ex]
&$2$&1.201907981039611&1.2015875&1.20158750&1.257083200326915&7850906020.699599&99344.05903710042\\[1ex]
&$3$&1.203404887789819&1.2030832&1.20308320&1.257158190890989&7870423087.973033&99470.43204300891\\[1ex]
&$4$&1.205397988323746&1.2050774&1.20507740&1.257258171002197&7896442041.004807&99638.67862380833\\[1ex]
&$5$&1.207892277880820&1.2075699&1.20756990&1.257383134399858&7928959622.092161&99848.58587960149\\[1ex]\hline
\multicolumn{7}{c}{ \bf CH } \\ \hline
$0$&$0$&0.1686796335222418&0.1686344&0.16863440&1.280743635551838&146502414.5533975&13697.88432686745\\[1ex]
$1$&$0$&0.5051421327711481&0.5050072&0.50500718&1.334184611139168&437954691.1786194&24172.55487834748\\[1ex]
&$1$&0.5087256719629778&0.5085903&0.50859034&1.334753837042328&444161731.2395528&24348.44091598796\\[1ex]
$2$&$0$&0.8416046320200543&0.841380&0.84137996&1.387625586726497&729406967.8038429&31814.20707264606\\[1ex]
&$1$&0.8451881712118841&0.8449631&0.84496312&1.388194812629657&735614007.8647747&31955.83749200750\\[1ex]
&$2$&0.8523507573365272&0.8521246&0.85212458&1.389332490341191&748011217.9336663&32237.18797002816\\[1ex]
$4$&$0$&1.514529630517867&1.5141255&1.51412550&1.494507537901155&1312311521.054287&44286.10911211507\\[1ex]
&$1$&1.518113169709697&1.5177087&1.51770870&1.495076763804316&1318518561.115218&44399.17187705266\\[1ex]
&$2$&1.525275755834340&1.5248701&1.52487010&1.496214441515850&1330915771.184110&44624.38119780340\\[1ex]
&$3$&1.536007762622479&1.5356002&1.53560020&1.497919028097772&1349469582.659151&44959.93956517626\\[1ex]
&$4$&1.550295071545778&1.5498843&1.54988430&1.500188222189207&1374130067.794166&45403.23494488982\\[1ex]
$5$&$0$&1.850992129766773&1.8504983&1.85049830&1.547948513488484&1603763797.679508&49825.13207814547\\[1ex]
&$1$&1.854575668958603&1.8540815&1.85408150&1.548517739391645&1609970837.740440&49930.63590766994\\[1ex]
&$2$&1.861738255083246&1.8612429&1.86124290&1.549655417103179&1622368047.809332&50140.91576569818\\[1ex]
&$3$&1.872470261871385&1.8719729&1.87197290&1.551360003685101&1640921859.284373&50454.53935639852\\[1ex]
&$4$&1.886757570794684&1.8862571&1.88625710&1.553629197776536&1665582344.419388&50869.41479505201\\[1ex]
&$5$&1.904579645811928&1.9040761&1.90407610&1.556459955019357&1696283716.548377&51382.85392189649\\[1ex]
\hline\hline
\end{tabular}\label{tab3} }
\vspace*{-13pt}
\end{table}

\begin{table}[!hbp]
\caption{The energy eigenvalues $E_{n, \ell}$(in eV), the expectation values  $\langle r^{2} \rangle $ (in $ (A^{o})^{2}$), $ \langle p^{2} \rangle $ (in $(eV/c)^{2}$) and the Heisenberg Uncertainty Relations HUR (in $ eV A^{o}/c$) corresponding to the pseudoharmonic potential for various $n$ and $\ell$ quantum numbers for $O_{2}$, $I_{2}$, LiH and HCl diatomic molecules.  \vspace*{13pt}} {\tiny
\begin{tabular}{c c c c c c}\hline
{}&{} &{} &{}&{} &{}\\[-1.5ex]
$n$&$\ell$&$E_{n, \ell}$ [FM] & $\langle r^{2} \rangle$ & $\langle p^{2} \rangle$ & HUR\\[2ex]\hline
\multicolumn{6}{c}{ \bf $O_{2}$ } \\ \hline
$0$&$0$&6.082405894024312E-02& 1.467870086578129& 453449135.0159387& 25799.31047672832\\[1ex]
$1$&$0$&0.1823830803004576&1.485069856601488&1359013219.817498&44924.70998764375\\[1ex]
&$1$&0.1827419925110476& 1.485120537234774&1364349752.232202&45013.59613507033\\[1ex]
$2$&$0$&0.3039421016606738&1.502269626624847&2264577304.619040&58326.71516443513\\[1ex]
&$1$&0.3043010138712638&1.502320307258132&2269913837.033761&58396.38390433131\\[1ex]
&$2$&0.3050169834489775&1.502421663244821&2280585790.008812&58535.47211561160\\[1ex]
$3$&$0$&0.4255011230208900&1.519469396648205&3170141389.420599&69404.12685332494\\[1ex]
&$1$&0.4258600352314801&1.519520077281491&3175477921.835303&69463.67725072471\\[1ex]
&$2$&0.4265760048091938&1.519621433268180&3186149874.810354&69582.62455072057\\[1ex]
&$3$&0.4276499591757634&1.519773454051252&3202155025.505520&69760.66372620103\\[1ex]
$4$&$0$&0.5470601443811045&1.536669166671564&4075705474.222158&79139.18709888101\\[1ex]
$5$&$0$&0.6686191657413207&1.553868936694923&4981269559.023701&87978.63395774535\\[1ex]
\hline
\multicolumn{6}{c}{ \bf $I_{2}$}  \\ \hline
$0$&$0$&5.424499700006802E-03&7.098394036762265&320680189.5596539&47710.73616365649\\[1ex]
$1$&$0$&1.627098533780424E-02&7.122689533744640&961765820.0079417&82766.89761060214\\[1ex]
&$1$&1.628025357504326E-02&7.122710358250389&962864811.0585099&82814.29323082611\\[1ex]
$2$&$0$&2.711747097560124E-02&7.146985030727015&1602851450.456229&107030.6279664366\\[1ex]
&$1$&2.712673921284026E-02&7.147005855232764&1603950441.506798&107067.4703023858\\[1ex]
&$2$&2.714527568731917E-02&7.147047504060674&1606148404.230512&107141.1169607948\\[1ex]
$3$&$0$&3.796395661339824E-02&7.171280527709389&2243937080.904476&126853.8619581419\\[1ex]
&$1$&3.797322485063725E-02&7.171301352215139&2245036071.955085&126885.1063702248\\[1ex]
&$2$&3.799176132511617E-02&7.171343001043049&2247234034.678800&126947.5721166007\\[1ex]
&$3$&3.801956603683410E-02&7.171405473825944&2250530930.321917&127041.2131267849\\[1ex]
$4$&$0$&4.881044225119524E-02&7.195576024691764&2885022711.352764&144081.2279670782\\[1ex]
$5$&$0$&5.965692788899224E-02&7.219871521674139&3526108341.801052&159555.7871069145\\[1ex]\hline
\multicolumn{6}{c}{ \bf $LiH$ } \\ \hline
$0$&$0$&9.710845082063990E-02&2.595085655742136&79803360.75277831&14390.84975842462\\[1ex]
$1$&$0$&0.2908587779619696&2.693142155955575&238645395.3562673&25351.64638753563\\[1ex]
&$1$&0.2927238938926831&2.694086095315485&241702867.6682745&25518.00021520058\\[1ex]
$2$&$0$&0.4846091051032984&2.791198656169013&397487429.9597564&33308.65323527427\\[1ex]
&$1$&0.4864742210340127&2.792142595528923&400544902.2717635&33442.16624345619\\[1ex]
&$2$&0.4902022355989111&2.794029425531101 &406653057.1155632&33707.57492853888\\[1ex]
$3$&$0$&0.6783594322446271&2.889255156382451&556329464.5632454&40092.11610949021\\[1ex]
&$1$&0.6802245481753415&2.890199095742362&559386936.8752526&40208.70078884599\\[1ex]
&$2$&0.6839525627402399& 2.892085925744539&565495091.7190522&40440.82585504762\\[1ex]
&$3$&0.6895397804451084&2.894913554768570&574640405.9599305&40786.44750810148\\[1ex]
$4$&$0$&0.8721097593859559&2.987311656595890&715171499.1667345&46221.64164031762\\[1ex]
$5$&$0$&1.065860086527285&3.085368156809328&874013533.7702235&51929.31277915241\\[1ex]
\hline
\multicolumn{6}{c}{ \bf $HCl$}  \\ \hline
$0$&$0$&0.1560617311494585&1.652050178604493& 142777592.1598772&15358.24686051258\\[1ex]
$1$&$0$& 0.4675281318814175&1.706824648069883&427134412.9930121& 27000.80636083549\\[1ex]
&$1$& 0.4701527552915152&1.707286254022739&431926335.1967491&27155.51315721777\\[1ex]
$2$&$0$& 0.7789945326133765&1.761599117535273&711491233.8261490&35402.85764796715\\[1ex]
&$1$&0.7816191560234742&1.762060723488129&716283156.0298859&35526.53115822570\\[1ex]
&$2$& 0.7868660265436169&1.762983542268707&725858842.0084449&35772.57598316216\\[1ex]
$3$&$0$& 1.090460933345335&1.816373587000663&995848054.6592839&42530.36683534621\\[1ex]
&$1$& 1.093085556755433&1.816835192953519&1000639976.863021&42637.98688307097\\[1ex]
&$2$& 1.098332427275576&1.817758011734097&1010215662.841582&42852.39333700694\\[1ex]
&$3$& 1.106199168605709&1.819141258763775&1024558847.741779&43171.95005971520\\[1ex]
$4$&$0$& 1.401927334077294&1.871148056466053&1280204875.492421&48943.36384696100\\[1ex]
$5$&$0$& 1.713393734809253&1.925922525931443& 1564561696.325556& 54892.84665749171\\[1ex]
\hline\hline
\end{tabular}\label{tab4} }
\vspace*{-13pt}
\end{table}

\begin{table}[!hbp]
\caption{The energy eigenvalues $E_{n, \ell}$(in eV), the expectation values  $\langle r^{2} \rangle $ (in $ (A^{o})^{2}$), $ \langle p^{2} \rangle $ (in $(eV/c)^{2}$) and the Heisenberg Uncertainty Relations HUR (in $ eV A^{o}/c$) corresponding to the pseudoharmonic potential for various $n$ and $\ell$ quantum numbers for $H_{2}$, ScH, TiH and VH diatomic molecules.  \vspace*{13pt}} {\tiny
\begin{tabular}{c c c c c c}\hline
{}&{} &{} &{}&{} &{}\\[-1.5ex]
$n$&$\ell$&$E_{n, \ell}$ [FM] & $\langle r^{2} \rangle$ & $\langle p^{2} \rangle$ & HUR\\[2ex]\hline
\multicolumn{6}{c}{ \bf $H_{2}$ } \\ \hline
$0$&$0$&0.3802143254317158&0.5720067969319436&179353151.5582044& 10128.73248449469\\[1ex]
$1$&$0$& 1.136872065928291&0.6158608064535510&534520083.1548633&18143.59306965887\\[1ex]
&$1$& 1.151940622653841&0.6167341564193854&548652322.0868371&18394.90763852220\\[1ex]
$2$&$0$& 1.893529806424867&0.6597148159751585&889687014.7515233&24226.83853110614\\[1ex]
&$1$& 1.908598363150416&0.6605881659409929&903819253.6834971&24434.65373466418\\[1ex]
&$2$& 1.938664759306279&0.6623307257790985&931950074.8372046&24844.70103375790\\[1ex]
$3$&$0$& 2.650187546921444&0.7035688254967659&1244853946.348182&29594.60134800272\\[1ex]
&$1$& 2.665256103646993&0.7044421754626004&1258986185.280156&29780.58037104236\\[1ex]
&$2$& 2.695322499802856&0.7061847353007059&1287117006.433864&30148.67132212521\\[1ex]
&$3$& 2.740245300798563&0.7087883221856139&1328983795.005013&30691.50035878782\\[1ex]
$4$&$0$& 3.406845287418019&0.7474228350183734&1600020877.944841&34581.67348006918\\[1ex]
$5$&$0$& 4.163503027914595&0.7912768445399808&1955187809.541501&39333.12650193264\\[1ex]
\hline
\multicolumn{6}{c}{ \bf $ScH$}  \\ \hline
$0$&$0$&7.793888021911055E-02& 3.208805616306735&71740403.45470674&15172.37652846693\\[1ex]
$1$&$0$& 0.2334806033879131& 3.317829166958441&214603976.5716611&26683.69038972540\\[1ex]
&$1$& 0.2348245288753450&3.318771086193946&217072082.5705855&26840.50206786684\\[1ex]
$2$&$0$& 0.3890223265567156& 3.426852717610148&357467549.6886156& 34999.83777259364\\[1ex]
&$1$& 0.3903662520441475& 3.427794636845652& 359935655.6875399& 35125.28306185265\\[1ex]
&$2$& 0.3930526196575466& 3.429677632258357& 364867451.6219697& 35374.81784500067\\[1ex]
$3$&$0$& 0.5445640497255191&3.535876268261853&500331122.8055699&42060.77678076121\\[1ex]
&$1$& 0.5459079752129510& 3.536818187497358&502799228.8044953&42170.00660535145\\[1ex]
&$2$& 0.5485940461540562&3.538701182910063&507731024.7389241&42387.59698123697\\[1ex]
&$3$& 0.5526210758596637& 3.541523572150610&515117708.0628493& 42711.84268486642\\[1ex]
$4$&$0$& 0.7001057728943216&3.644899818913559&643194695.9225243&48418.80038470771\\[1ex]
$5$&$0$& 0.8556474960631242& 3.753923369565265& 786058269.0394796& 54321.28962006814\\[1ex]
\hline
\multicolumn{6}{c}{ \bf $TiH$ } \\ \hline
$0$&$0$&7.414243861897418E-02& 3.229321318744519& 68344509.11468062& 14856.18996590873\\[1ex]
$1$&$0$&0.2220937097705074& 3.343783679169409&204419756.4028107& 26144.51080359900\\[1ex]
&$1$& 0.2234282082928383& 3.344816082070214& 206873941.5371264& 26305.04298826812\\[1ex]
$2$&$0$& 0.3700449809220414& 3.458246039594299& 340495003.6909408& 34314.94569449065\\[1ex]
&$1$& 0.3713794794443714& 3.459278442495104& 342949188.8252565& 34443.52966486439\\[1ex]
&$2$& 0.3740470655157591& 3.461342241248182&347852773.6980163& 34699.24349804025\\[1ex]
$3$&$0$& 0.5179962520735746& 3.572708400019190& 476570250.9790701& 41263.13777298301\\[1ex]
&$1$& 0.5193307505959055& 3.573740802919994& 479024436.1133866& 41375.22414361222\\[1ex]
&$2$& 0.5219983366672931& 3.575804601673072& 483928020.9861464& 41598.46204273429\\[1ex]
&$3$& 0.5259964705358442& 3.578897787084974& 491271469.7330276& 41931.01925645876\\[1ex]
$4$&$0$& 0.6659478054197630& 3.687170760444080& 612645498.2672009& 47528.18708649129\\[1ex]
$5$&$0$& 0.8138990765712970& 3.801633120868971& 748720745.5553311& 53351.30349471187\\[1ex]
\hline
\multicolumn{6}{c}{ \bf $VH$}  \\ \hline
$0$&$0$& 8.186329543168380E-02& 3.006871384646724& 75505142.31907943& 15067.65581744282\\[1ex]
$1$&$0$& 0.2452318258208468&3.110465298102923& 225856581.7073171& 26505.07611286482\\[1ex]
&$1$& 0.2466633196345676& 3.111373062957315& 228491052.2331402& 26663.10006441421\\[1ex]
$2$&$0$& 0.4086003562100107& 3.214059211559122& 376208021.0955538& 34772.90404416337\\[1ex]
&$1$& 0.4100318500237314& 3.214966976413514& 378842491.6213778& 34899.37105199667\\[1ex]
&$2$& 0.4128935912481992& 3.216781670327672& 384106583.7945811& 35150.91774339026\\[1ex]
$3$&$0$& 0.5719685749984311& 3.317653125015321& 526559460.4837915& 41796.43094308930\\[1ex]
&$1$& 0.5734006920136379& 3.318560889869713& 529193931.0096155& 41906.58996631581\\[1ex]
&$2$& 0.5762624332381066& 3.320375583783870& 534458023.1828187& 42126.01774122051\\[1ex]
&$3$& 0.5805513058658915& 3.323095539010299& 542342072.6561637& 42452.96835630339\\[1ex]
$4$&$0$& 0.7353371053875950& 3.421247038471519& 676910899.8720292& 48123.58477395747\\[1ex]
$5$&$0$& 0.8987056357767589& 3.524840951927718& 827262339.2602668& 53999.70529004866\\[1ex]
\hline\hline
\end{tabular}\label{tab5} }
\vspace*{-13pt}
\end{table}

\begin{table}[!hbp]
\caption{The energy eigenvalues $E_{n, \ell}$(in eV), the expectation values  $\langle r^{2} \rangle $ (in $ (A^{o})^{2}$), $ \langle p^{2} \rangle $ (in $(eV/c)^{2}$) and the Heisenberg Uncertainty Relations HUR (in $ eV A^{o}/c$) corresponding to the pseudoharmonic potential for various $n$ and $\ell$ quantum numbers for CrH, MnH, CuLi and TiC diatomic molecules.  \vspace*{13pt}} {\tiny
\begin{tabular}{c c c c c c}\hline
{}&{} &{} &{}&{} &{}\\[-1.5ex]
$n$&$\ell$&$E_{n, \ell}$ [FM] & $\langle r^{2} \rangle$ & $\langle p^{2} \rangle$ & HUR\\[2ex]\hline
\multicolumn{6}{c}{ \bf $CrH$ } \\ \hline
$0$&$0$& 7.939774230820884E-02&2.923120140335614& 73312725.17284806& 14639.05405057451\\[1ex]
$1$&$0$& 0.2378248211383456& 3.029840333495707& 219259743.6216898& 25774.44499377094\\[1ex]
&$1$& 0.2392973231890112& 3.030832335554967& 221972416.2499066& 25937.64015425235\\[1ex]
$2$&$0$& 0.3962518999684832& 3.136560526655800& 365206762.0705316& 33845.13427330738\\[1ex]
&$1$& 0.3977244020191479& 3.137552528715060& 367919434.6987484& 33975.97022459364\\[1ex]
&$2$& 0.4006681974177901& 3.139535505170499& 373339162.6082527& 34236.11479825407\\[1ex]
$3$&$0$& 0.5546789787986199& 3.243280719815893& 511153780.5193725& 40716.27685851796\\[1ex]
&$1$& 0.5561514808492856& 3.244272721875153&513866453.1475893& 40830.41656208592\\[1ex]
&$2$& 0.5590952762479269& 3.246255698330592& 519286181.0570936& 41057.71211746852\\[1ex]
&$3$& 0.5635070410621514& 3.249227599169900& 527401773.0905892& 41396.23650740588\\[1ex]
$4$&$0$& 0.7131060576287567& 3.350000912975986& 657100798.9682143& 46917.88866158374\\[1ex]
$5$&$0$& 0.8715331364588943& 3.456721106136079& 803047817.4170561& 52686.92759785912\\[1ex]
\hline
\multicolumn{6}{c}{ \bf $MnH$}  \\ \hline
$0$&$0$& 6.791703912318869E-02& 3.135496864142230& 62788898.48101475& 14031.19361601737\\[1ex]
$1$&$0$& 0.2034073084815335& 3.260156395169919& 187733167.6563940& 24739.43182695390\\[1ex]
&$1$& 0.2047808489433400& 3.261420303318240& 190266106.7993997& 24910.59500993260\\[1ex]
$2$&$0$& 0.3388975778398784& 3.384815926197607& 312677436.8317725& 32532.37722563215\\[1ex]
&$1$& 0.3402711183016849& 3.386079834345928& 315210375.9747782& 32669.97853786860\\[1ex]
&$2$& 0.3430169070893978& 3.388606093133057& 320270015.6021789& 32943.42007620585\\[1ex]
$3$&$0$& 0.4743878471982228& 3.509475457225295& 437621706.0071517& 39189.57306199141\\[1ex]
&$1$& 0.4757613876600293& 3.510739365373616& 440154645.1501575& 39309.90001997860\\[1ex]
&$2$& 0.4785071764477422& 3.513265624160745& 445214284.7775581& 39549.41266434063\\[1ex]
&$3$& 0.4826215955808415& 3.517051128144562& 452788205.0614842& 39905.88011085639\\[1ex]
$4$&$0$& 0.6098781165565677& 3.634134988252983& 562565975.1825310& 45215.49174355506\\[1ex]
$5$&$0$& 0.7453683859149125& 3.758794519280671& 687510244.3579102& 50835.12307885000\\[1ex]
\hline
\multicolumn{6}{c}{ \bf $CuLi$ } \\ \hline
$0$&$0$& 2.088511890200140E-02& 5.368123991963938& 121864956.0741156& 25577.06383620081\\[1ex]
$1$&$0$& 6.262381995154742E-02& 5.432124528796563& 365230008.4756823& 44541.83300778416\\[1ex]
&$1$& 6.274896727143631E-02& 5.432316428163798& 366689388.4190958& 44631.52236639918\\[1ex]
$2$&$0$& 0.1043625210010930& 5.496125065629188& 608595060.8772444& 57835.23639534591\\[1ex]
&$1$& 0.1044876683209819& 5.496316964996423& 610054440.8206624& 57905.54872077471\\[1ex]
&$2$& 0.1047379629607597& 5.496700743029687& 612972885.8630584& 58045.91731707342\\[1ex]
$3$&$0$& 0.1461012220506386& 5.560125602461812& 851960113.2788110& 68825.90528367776\\[1ex]
&$1$& 0.1462263693705275&5.560317501829047& 853419493.2222290& 68886.01704675367\\[1ex]
&$2$& 0.1464766640103052& 5.560701279862312& 856337938.2646205& 69006.08284276634\\[1ex]
&$3$& 0.1468521059699719& 5.561276895170364& 860714818.9717922& 69185.78926397074\\[1ex]
$4$&$0$& 0.1878399231001846& 5.624126139294437& 1095325165.680378& 78487.24033452841\\[1ex]
$5$&$0$& 0.2295786241497302& 5.688126676127061& 1338690218.081940& 87261.90199933894\\[1ex]
\hline
\multicolumn{6}{c}{ \bf $TiC$}  \\ \hline
$0$&$0$& 2.689705655191421E-02& 3.220299274172334& 240826937.3569261& 27848.42565696767\\[1ex]
$1$&$0$& 8.065722427377420E-02& 3.252677646120945& 721873174.7872818& 48456.37975504308\\[1ex]
&$1$& 8.079298641607657E-02& 3.252759443059853& 724303654.1180329& 48538.49555327389\\[1ex]
$2$&$0$& 0.1344173919956333& 3.285056018069556& 1202919412.217638& 62862.21165659258\\[1ex]
&$1$& 0.1345531541379357& 3.285137815008464& 1205349891.548389& 62926.46906542558\\[1ex]
&$2$& 0.1348246784225422& 3.285301402622239& 1210210477.970561& 63054.78713582985\\[1ex]
$3$&$0$& 0.1881775597174924& 3.317434390018167& 1683965649.648004& 74742.52844098581\\[1ex]
&$1$& 0.1883133218597957& 3.317516186957075& 1686396128.978745& 74797.36930874464\\[1ex]
&$2$&0.1885848461444013& 3.317679774570850& 1691256715.400927& 74906.93024208630\\[1ex]
&$3$& 0.1889925427288395& 3.317925140333806& 1698546664.649718& 75070.97095862975\\[1ex]
$4$&$0$& 0.2419377274393515& 3.349812761966778& 2165011887.078360& 85160.93264604882\\[1ex]
$5$&$0$& 0.2956978951612106& 3.382191133915389& 2646058124.508715& 94601.66134132190\\[1ex]
\hline\hline
\end{tabular}\label{tab6} }
\vspace*{-13pt}
\end{table}

\begin{table}[!hbp]
\caption{The energy eigenvalues $E_{n, \ell}$(in eV), the expectation values  $\langle r^{2} \rangle $ (in $ (A^{o})^{2}$), $ \langle P^{2} \rangle $ (in $(eV/c)^{2}$) and the Heisenberg Uncertainty Relations HUR (in $ eV A^{o}/c$) corresponding to the pseudoharmonic potential for various $n$ and $\ell$ quantum numbers for NiC, ScN, ScF and $Ar_{2}$ diatomic molecules.  \vspace*{13pt}} {\tiny
\begin{tabular}{c c c c c c}\hline
{}&{} &{} &{}&{} &{}\\[-1.5ex]
$n$&$\ell$&$E_{n, \ell}$ [FM] & $\langle r^{2} \rangle$ & $\langle P^{2} \rangle$ & HUR\\[2ex]\hline
\multicolumn{6}{c}{ \bf NiC } \\ \hline
$0$&$0$& 2.969207342798175E-02& 2.641775184923072& 276052154.6005498& 27004.95754057539\\[1ex]
$1$&$0$& 8.903609107369803E-02& 2.670024239769140& 827415521.7954127& 47002.33504364418\\[1ex]
&$1$& 8.919546221483721E-02& 2.670100162872615& 830379193.4841548& 47087.10672538792\\[1ex]
$2$&$0$& 0.1483801087194152& 2.698273294615207& 1378778888.990264& 60994.44446293167\\[1ex]
&$1$&0.1485394798605535& 2.698349217718682& 1381742560.679018& 61060.82179021862\\[1ex]
&$2$& 0.1488586749017538& 2.698501057345057& 1387669390.331840& 61193.36007326155\\[1ex]
$3$&$0$& 0.2077241263651315& 2.726522349461275& 1930142256.185127& 72543.61446142835\\[1ex]
&$1$& 0.2078834975062698& 2.726598272564750& 1933105927.873869& 72600.29809598284\\[1ex]
&$2$& 0.2082026925474700& 2.726750112191124& 1939032757.526702& 72713.53236590972\\[1ex]
&$3$& 0.2086808059708858& 2.726977855182097& 1947921718.028147& 72883.05282225095\\[1ex]
$4$&$0$& 0.2670681440108478& 2.754771404307343& 2481505623.379979& 82679.98990635540\\[1ex]
$5$&$0$&0.3264121616565641& 2.783020459153410& 3032868990.574830& 91872.39221170690\\[1ex]
\hline
\multicolumn{6}{c}{ \bf ScN}  \\ \hline
$0$&$0$& 3.380444182356435E-02& 3.137410325933636& 336542668.3090815& 32494.19090653237\\[1ex]
$1$&$0$& 0.1013822471073453& 3.160572203976006& 1009005150.475157& 56471.52939544323\\[1ex]
&$1$& 0.1015080480754342&3.160615110101562& 1011496529.811182& 56541.58837472217\\[1ex]
$2$&$0$& 0.1689600523911281& 3.183734082018375& 1681467632.641253& 73166.56210217006\\[1ex]
&$1$& 0.1690858533592170& 3.183776988143932& 1683959011.977278& 73221.23975535276\\[1ex]
&$2$& 0.1693353929844399& 3.183862798628290& 1688941565.477345& 73330.47265209974\\[1ex]
$3$&$0$& 0.2365378576749109& 3.206895960060745& 2353930114.807349& 86883.88213840939\\[1ex]
&$1$& 0.2366636586429980& 3.206938866186301& 2356421494.143354& 86930.43008455163\\[1ex]
&$2$& 0.2369131982682227& 3.207024676670659& 2361404047.643441& 87023.45116336453\\[1ex]
&$3$& 0.2372895700170101& 3.207153387980668& 2368877365.027091& 87162.79634659131\\[1ex]
$4$&$0$& 0.3041156629586919& 3.230057838103114& 3026392596.973445& 98870.73949875825\\[1ex]
$5$&$0$& 0.3716934682424746& 3.253219716145483& 3698855079.139521& 109695.8899440701\\[1ex]
\hline
\multicolumn{6}{c}{ \bf ScF } \\ \hline
$0$&$0$& 3.373955790469907E-02& 3.227717235471636& 420001022.2062209& 36819.07845520750\\[1ex]
$1$&$0$& 0.1011947411226970& 3.246272807998829& 1259398134.700247& 63940.20581075619\\[1ex]
&$1$& 0.1012925231418258& 3.246299552935829& 1261817839.752488& 64001.86473123229\\[1ex]
$2$&$0$& 0.1686499243406931& 3.264828380526023& 2098795247.194306& 82778.05438613001\\[1ex]
&$1$& 0.1687477063598237& 3.264855125463022& 2101214952.246547& 82826.09737601837\\[1ex]
&$2$& 0.1689422411136725& 3.264908614670295& 2106054241.710537& 82922.09980893905\\[1ex]
$3$&$0$& 0.2361051075586893& 3.283383953053217& 2938192359.688364& 98220.23032341320\\[1ex]
&$1$& 0.2362028895778199& 3.283410697990216& 2940612064.740605& 98261.06610457877\\[1ex]
&$2$& 0.2363974243316687& 3.283464187197489& 2945451354.204563& 98342.68674722603\\[1ex]
&$3$& 0.2366887118202357& 3.283544419341662& 2952709986.821642& 98464.99072849496\\[1ex]
$4$&$0$& 0.3035602907766872& 3.301939525580410& 3777589472.182390& 111684.2513052555\\[1ex]
$5$&$0$& 0.3710154739946834& 3.320495098107603& 4616986584.676449& 123817.1285503210\\[1ex]
\hline
\multicolumn{6}{c}{ \bf $Ar_{2}$ } \\ \hline
$0$&$0$& 6.364744498285280E-03& 6.413083011982251& 319839318.8285583& 45289.69090359750\\[1ex]
$1$&$0$& 1.909132682864900E-02& 6.437443531035714& 959213790.5225470& 78580.43401941488\\[1ex]
&$1$&1.910336674723512E-02& 6.437466708774742& 960430449.4184439& 78630.39516767542\\[1ex]
$2$&$0$& 3.181790915901272E-02& 6.461804050089178& 1598588262.216536& 101635.4470999948\\[1ex]
&$1$& 3.182994907759884E-02& 6.461827227828205& 1599804921.112433& 101674.2986130609\\[1ex]
&$2$& 3.185441729924232E-02& 6.461873583054483& 1602238212.471537& 101751.9571257981\\[1ex]
$3$&$0$& 4.454449148937645E-02& 6.486164569142642& 2237962733.910562& 120481.5114104745\\[1ex]
&$1$& 4.455653140796256E-02& 6.486187746881669& 2239179392.806459& 120514.4719139207\\[1ex]
&$2$& 4.458099962960604E-02& 6.486234102107946& 2241612684.165525& 120580.3658799896\\[1ex]
&$3$& 4.461711938536572E-02& 6.486303634317930& 2245262555.124323& 120679.1393377532\\[1ex]
$4$&$0$& 5.727107381974017E-02& 6.510525088196106& 2877337205.604551& 136868.4626358041\\[1ex]
$5$&$0$& 6.999765615010389E-02& 6.534885607249569& 3516711677.298540&151595.8723871621\\[1ex]
\hline\hline
\end{tabular}\label{tab7} }
\vspace*{-13pt}
\end{table}

\section{Conclusions}
We have used $SU(1,1)$ spectrum generating algebra approach to
obtain the solutions of the $3$-dimensional Schr\"{o}dinger wave
equation with  pseudoharmonic molecular potential. The explicit
bound state energies, the eigenfunctions and the radial matrix
elements are obtained for this molecular potential. Furthermore,
based upon the solutions obtained, by using Hellmann-Feynman
theorem, the expectation values of $r^{2}$ and $p^{2}$ are
obtained. The Heisenberg uncertainty products $P_{n, \ell}$ for
this potential are obtained also.

The solutions obtained have been applied to compute the explicit
bound state energies for some selected diatomic molecules. The
selected diatomic molecules composed of the some homogeneous
diatomic molecules (dimers) ($O_{2}, I_{2}, N_{2}, H_{2}$ and $Ar_{2}$); the
heterogeneous diatomic molecules (CO, NO, HCl, CH and LiH); the
neutral transition metal hydrides ( ScH, TiH, VH, CrH and MnH); the
transition-metal lithide (CuLi); the transition-metal carbides
(TiC and NiC); the transition-metal nitrite (ScN) and the
transition-metal fluoride(ScF).

The explicit bound state energies obtained for these diatomic
molecules for various values of $n$ and $\ell$ are compared with
the exact method \cite{IkS07} and the Nikiforov-Uvarov method
\cite{SeE071} for CO, NO, CH and $ N_{2}$ and our results are in
excellent agreement with their  results as displayed in Tables 2 - 7.

Though, slight differences are noticed in the explicit bound state energies we obtained
when compared with the results of Ikhdair and Sever and Sever et al. \cite{IkS07, SeE071} ,
this is due to the conversions used by Ikhdair and Sever and Sever et al. \cite{IkS07, SeE071}
as cited in the work of Ikhdair \cite{Ikh09}, they have used the following
conversions: 1amu = 931.502 MeV/$c^{2}$, 1$c m^{-1}$ = 1.23985 X $10^{-4}$ eV,
and $\hbar$ c = 1973.29~ eV $A^{o}$ (cf. pp. 791, Bransden and Joachain (2000) \cite{BrJ00}).
In our calculations, we have used the following recent conversions:
1amu = 931.494 028 MeV/$c^{2}$, 1$c m^{-1}$ = 1.239841875 X $10^{-4}$
eV, and $\hbar$ c = 1973.29~ eV $A^{o}$ (cf. pp. 4, Nakamura et al. (2010) in the 2010
edition of the Review of Particle Physics\cite{NaE10}). This explains the reason for the slight differences.

In addition, we obtained the numerical values of the expectation
values of $r^{2}$ and $p^{2}$, and the Heisenberg uncertainty
product for these diatomic molecules. These results are displayed
together with the explicit bound state energies of these diatomic
molecules for various values of $n$ and $\ell$ in Tables 2 - 7.
The Heisenberg uncertainty products obtained are valid in each
case of these  diatomic molecules for various $n$ and $\ell$, as
expected from equation (\ref{p49}) that
\begin{equation}
P_{n, \ell}\geq  2959.89~ eV A^{o}/c. 
\label{p51}
\end{equation}

This implies that the numerical value of the Heisenberg
uncertainty product $P_{n, \ell}$ can not be less than $2959.89 ~eV
A^{o}/c$ for this principle to hold. The Heisenberg uncertainty
products in all these selected diatomic molecules attains its
minimum value of $2959.89 ~eV A^{o}/c$, even the lowest Heisenberg
uncertainty products (for ground state)  $P_{0,0}$ obtained for
$H_{2}$ is $10128.73248449469~ eV A^{o}/c$ which is greater than
the minimum.

It is evident from the Tables displayed that the explicit bound
state energies, the expectation values of $r^{2}$ and $p^{2}$ and
the Heisenberg uncertainty products $P_{n, \ell}$ increase as the
quantum numbers $(n, \ell)$ of the state increase. Similar studies
involving the confined diatomic molecules are currently in
progress \cite{OyE11}.

It should be noted that the advantage of this SGA method is that,
it allows one to find the explicit bound state energies and the
eigenfunctions directly in  a simple and unique way. This method,
as applied here to the diatomic molecules demonstrates that the
values obtained are in excellent agreement with earlier results
derived from the other methods. We have also demonstrated that the
Heisenberg uncertainty product is validated by all the diatoms
considered. Finally, the method serves as a very useful link for
finding the matrix elements from the creation and annihilation
operators in addition to allowing the construction of the coherent
states.

[Note to the desk Editor: Please, put the figure in the section $2$,
where it is indicated in the text]

\vspace{5mm}
{\bf  \huge{Acknowledgements}.}

One of the authors KJO thanks the TWAS-UNESCO Associateship Scheme at Centres
of Excellence in the South, Trieste, Italy for the award and
the financial support. He also thanks his
host Institution, School of Chemistry, University of Hyderabad, India.
He acknowledges University of Ilorin for granting him leave. 
KJO ackonowledges eJDS (ICTP), Profs. P. G. Hajiogeorgiou, G. Van Hooydonk, Z. Rong and D. Popov for communicating some of their research materials to him.
Finally, the authors also wish to thank the School of Chemistry, University of Hyderabad. \\

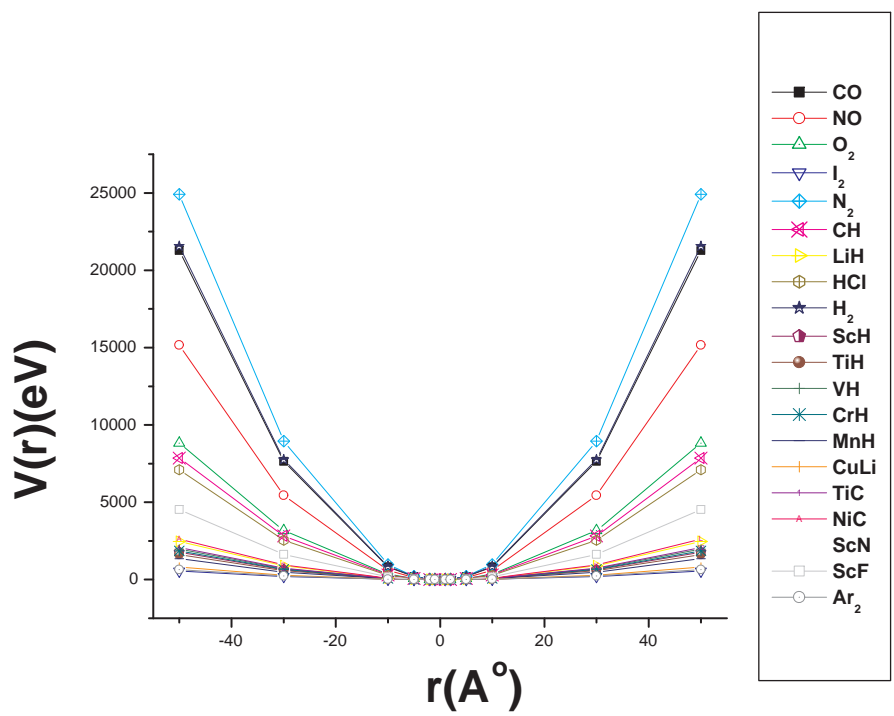
\end{document}